\documentclass[11pt,twoside,usenatbib]{article}
\usepackage{asp2014}

\aspSuppressVolSlug
\resetcounters

\begin{document}

\title{3D maps of the Magellanic Clouds using Classical Cepheids}
\author{Abigail H. Chown, Victoria Scowcroft
\affil{Department of Physics, University of Bath, Bath, BA2 7AY, United Kingdom; \email{A.H.Chown@bath.ac.uk}}}

\paperauthor{Abigail H. Chown}{A.H.Chown@bath.ac.uk}{}{University of Bath}{Department of Physics}{Bath}{Somerset}{BA2 7AY}{United Kingdom}
\paperauthor{Victoria Scowcroft}{V.Scowcroft@bath.ac.uk}{}{University of Bath}{Department of Physics}{Bath}{Somerset}{BA2 7AY}{United Kingdom}

\begin{abstract}
Variable stars have been used for over one hundred years as probes for determining astronomical distances; these distances can be used to map the three-dimensional (3D) structure of nearby galaxies. Exploiting the effect that moving to the mid-infrared has on Cepheid magnitudes and light curves, we can now map our nearest galaxies in 3D at fidelities never before achievable. Combining data from the OGLE-IV catalogue with mid-infrared photometry from the \textit{Spitzer Space Telescope}, $\sim$5000 fundamental mode Cepheids are being used to trace the 3D structure of the Magellanic Clouds. An automated photometry pipeline has been developed to obtain precise mean magnitudes and light curves for Cepheids in the Magellanic System, which in turn produces distance measurements for individual Cepheids accurate to 5\%. The resulting detailed maps are being used to probe the geometric and chemical structure of the Magellanic Clouds, as well as their interaction and dynamical histories. Initial results from this project are discussed and the future inclusion of RR Lyrae stars to trace the old stellar population of the system is described. 
\end{abstract}

\section{Introduction}
First discovered by Henrietta Leavitt around a century ago, the Cepheid Leavitt Law (\citep{Leavitt, Leavitt1912}, previously known as the period-luminosity relation) is an important tool in astronomy for determining distances. We use the Leavitt Law (LL) to determine individual distances to Cepheids in nearby galaxies and, in turn, construct three-dimensional (3D) maps of the structure traced by these variable stars. Of particular interest are the Large and Small Magellanic Cloud (LMC and SMC, respectively), which are two dwarf galaxies in our Local Group that are interacting with each other and with the Milky Way.  

 Constructing a 3D map of a nearby galaxy has numerous advantages compared to just looking at the 2D sky projection. First, galaxies are 3D objects and we should therefore treat them this way by endeavouring to understand their complex and intrinsic 3D structure. For example, \cite{Scowcroft2016} used Cepheids to show that while the SMC appears small on the sky relative to the LMC, it is actually in fact extremely elongated along the line of sight by $\sim 20$ kpc. This was also shown by \cite{Jacyszyn-Dobrzeniecka2016a}, who also found that the Cepheids situated in the Magellanic Bridge (MB) - a stream of dust, gas and stars between the two Clouds - form a continuous-like connection between the two galaxies. 

Along with the geometric structure, 3D maps of galaxies can be used to study the chemical structure of the galaxy by looking for correlations between the metallicity and 3D position of Cepheids. For example, \cite{Ripepi2017} derived photometric metallicities from the $R_{31}$ Fourier parameter of the light curve of all SMC Cepheids with periods between 2.5 and 4 days. The metallicity distribution of these Cepheids was found to peak at [Fe/H] = -0.6 dex, with the more metal-rich Cepheids being situated closer to the centre of the galaxy.

As well as the metallicity, another parameter of interest for a Cepheid is its age. The Magellanic Bridge Cepheids are a crucial population for such a study as understanding their age distribution will be key for determining whether these Cepheids formed in-situ or if they were possibly ejected from one of the Clouds. Using the period-age relations of \cite{Bono2005} without stellar rotation and the relations of \cite{Anderson2016} with stellar rotation, \cite{Jacyszyn-Dobrzeniecka2019} determined the ages of 10  Magellanic Bridge Classical Cepheids. Without accounting for stellar rotation, the majority of the Cepheids were less than 300 Myr old, suggesting that they were formed in-situ. However, the age approximately doubles when stellar rotation is accounted for, suggesting that some of the MB Cepheids could have migrated from one of the Clouds. 

Our work studies the entire fundamental mode classical Cepheid population in the Magellanic Clouds and will produce 3D maps of the system, which will be used to study the system's geometric and chemical structure as well as its dynamical and evolutionary history.

\subsection{Why the mid-infrared?}
We work with data from the \textit{Spitzer Space Telescope}, which operates at mid-infrared wavelengths ($3.6$ $\mu$m and $4.5$ $\mu$m). There are several benefits of working at mid-infrared wavelengths. First, at longer wavelengths, the effects of interstellar extinction are substantially reduced. For example, the extinction in the \textit{Spitzer} $3.6$ $\mu$m band is around 20 times smaller than the extinction in the $V$ band \citep{Indebetouw2005}. Therefore, working at this wavelength regime allows for any determined magnitudes to carry smaller uncertainties as they are less affected by interstellar extinction. Second, the Cepheid light curves also benefit from working in the mid-infrared. As we move to longer wavelengths, the light curve becomes more sinusoidal and less saw-tooth like. In addition, the amplitude of the light curve decreases as the wavelength increases. A smaller amplitude is advantageous because fewer observations are required to obtain precise mean magnitudes as each individual observation is closer to mean light. Finally, the slope of the Leavitt Law becomes steeper for increasing wavelengths and the dispersion reduces. Therefore, magnitudes inferred from a mid-infrared LL will carry smaller uncertainties than those inferred from optical relations.

\section{Data}
Our work employs a template fitting method and therefore we work with two datasets - a calibrating sample and a complete sample of Cepheids. The calibrating sample is used to produce the templates, and the complete sample contains all known fundamental mode classical Cepheids to which the templates will be applied. 

The calibrating sample contains 85 LMC and 90 SMC Cepheids, which were observed by \textit{Spitzer} as part of the Carnegie Hubble Program (CHP) \citep{Freedman2011a, Scowcroft2011, Scowcroft2016} and were chosen as they are bright, isolated targets. For each LMC Cepheid there are 24 epochs of observation while each SMC Cepheid has 12 epochs. These observations were taken in such a way as to fully sample the entire light curve. Each epoch of observation has 5 FITS images in the $3.6 \mu m$ band and 5 in the $4.5 \mu m$ band. A medium-scale, five-dither pattern was used.  

The complete sample of Cepheids was taken from the OGLE-IV catalogue of variable stars \citep{Soszynski2015}. The OGLE-IV catalogue provides 99\% coverage of the Magellanic Clouds and currently contains 2476 LMC and 2753 SMC fundamental mode Classical Cepheids. The catalogue contains the period, $V$ and $I$ magnitudes, RA and Dec of each Cepheid. This catalogue was combined with mid-infrared observations taken as part of the \textit{Spitzer} legacy program SAGE (Surveying the Agents of Galaxy Evolution), which performed a uniform, unbiased survey of the Magellanic Clouds. From the SAGE survey, each Cepheid in the complete sample has a minimum of two and a maximum of six epochs of observation. This incomplete sampling of the infrared light curve from SAGE requires the use of template fitting to obtain mean magnitudes for these Cepheids.  

\section{Photometry pipeline for the calibrating sample}

In this section, we give a brief overview of the photometry pipeline that has been developed to obtain mid-infrared light curves and mean magnitudes for the calibrating sample. For each Cepheid, the pipeline carries out the following: 

 \textbf{\textit{\underline{Input files}}} For each Cepheid, there are five dithered images of the target in the $3.6$ $\mu$m band and five in the $4.5$ $\mu$m band, per epoch of observation. Therefore, there are 240 images for each LMC Cepheid and 120 for each SMC Cepheid. The pipeline combines all of these available images for the Cepheid into one medianed image that is used in subsequent steps of the pipeline. 
 
 \textbf{\textit{\underline{PSF photometry}}} Using the \verb|DAOPHOT/ALLSTAR/ALLFRAME| \citep{Stetson1987, Stetson1988, Stetson1994} the pipeline performs PSF photometry for all stars in the field for every individual FITS image. From the median image, a master star list and PSF model are created, as the median image has a much higher S/N ratio than a single image. \verb|ALLFRAME| is then used to obtain instrumental magnitudes. 

\begin{figure}[t!]
\centering
  \includegraphics[width=10cm]{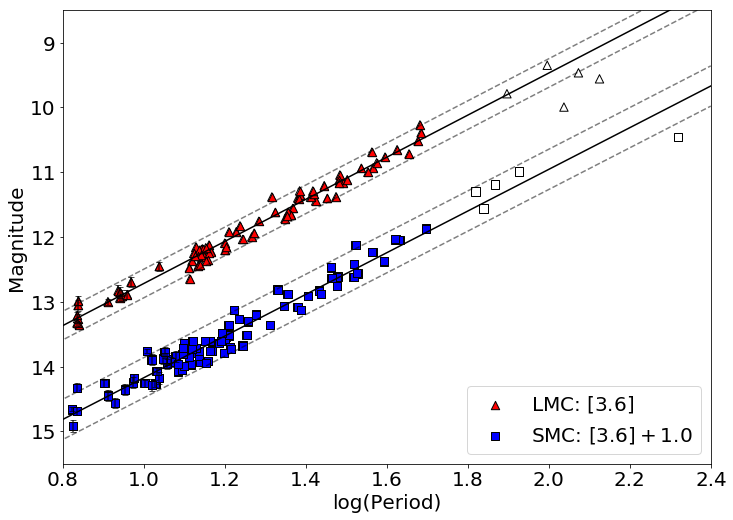}
  \caption{The mid-infrared $3.6$ $\mu$m Leavitt Law for the LMC (red triangles) and SMC (blue squares) calibrating Cepheids. The SMC LL relation is offset by a value of 1.0 in magnitude from the LMC LL relation for clarity. Only Cepheids with periods between 6 and 60 days were included in the fit due to the debate surrounding the nonlinearity of the PL relation at long periods. The solid line shows the best-fitting relation and the dashed lines show $\pm 2\sigma$ for each galaxy.}
  \label{fig:PL}
\end{figure}

\textbf{\textit{\underline{Calibration procedure}}} The instrumental magnitudes obtained from the PSF photometry are then calibrated to the standard IRAC Vega system \citep{Reach2005}. This is carried out by applying an aperture correction and putting the magnitudes on the correct zero point. A location correction and a pixel-phase response correction is then applied to the magnitudes. 

\textbf{\textit{\underline{Light curves}}} As there are five dithered images in each band per epoch, there are five magnitude measurements for each band per epoch; these are combined into a weighted mean which serves as the final mean magnitude for that epoch. The magnitudes are then phased using the periods from OGLE-IV and the data are fit using GLOESS \citep{Persson2004}, a Gaussian local regression smoother. From this, mid-infrared light curves, colour curves and mean magnitudes are obtained for each Cepheid. 

Figure \ref{fig:PL} shows the resulting $3.6$ $\mu$m Leavitt Law for the calibrating sample. The SMC LL relation (blue squares) is offset from the LMC LL relation (red triangles) for clarity. Only Cepheids with periods between 6 and 60 days were included in the fit due to the debate surrounding the nonlinearity of the PL relation at long periods \citep{Tammann2003, Ngeow2005}. The slope and intercept values are in agreement with the literature (e.g. with \citet{Scowcroft2011, Scowcroft2016}), as is the observed scatter ($\sigma_{LMC}=0.110$ and $\sigma_{SMC}=0.155$). 

\section{Template light curves}

\subsection{Template construction}

The light curves from the well-sampled set of calibrating Cepheids were used to produce template light curves. These templates can then be applied to the SAGE data, which has fewer observations. The light curves were normalised following the approach of \cite{Soszynski2005} so that they could be compiled together in appropriate bins. The phase zero point was anchored to the mean magnitude along the rising branch \citep{Inno2013} to ensure that all normalised light curves started at the same point in their cycle. After testing various different ways of binning these light curves - including by period, colour, galaxy, wavelength - we decided to bin the Cepheids by galaxy and wavelength as this resulted in the smallest mean residual. Therefore, each galaxy has two templates: one for each of the $3.6$ $\mu$m and $4.5$ $\mu$m bands. A full description of the template procedure will be presented in Chown et al. (2020, in prep).

\subsection{Template fitting}

\begin{figure}[t!]
    \centering
    \includegraphics[width=13cm]{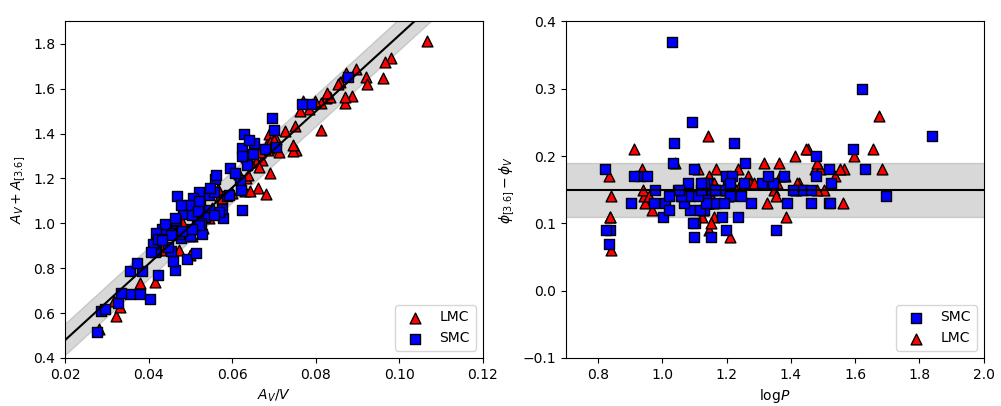}
    \caption{\textit{Left:} The amplitude relation for the calibrating sample allows the infrared $3.6$ $\mu$m amplitude to be inferred from the $V$ band amplitude and magnitude. \textit{Right:} The phase relation between the $3.6$ $\mu$m and $V$ bands for the calibrating sample. This relation allows the $3.6$ $\mu$m phase of the mean magnitude along the rising branch to be inferred from the $V$ band phase. In both panels, the LMC Cepheids are shown by red triangles and the SMC Cepheids are shown by blue squares. The best-fitting relation for each panel is shown by the solid line with $\pm 1\sigma$ shown by the shaded region. Similar results hold for the $4.5$ $\mu$m and $I$ band combinations.}
    \label{fig:relations}
\end{figure}

 To fit the templates to the observed SAGE data, the templates had to be scaled and shifted in amplitude and phase, respectively. As there were too few observations in the SAGE data to infer this directly, relations were found between the Cepheid parameters at optical and infrared wavelengths and are shown in Figure \ref{fig:relations}. Rather than assuming that the infrared amplitude and phase obtained from the relations are completely accurate, Monte Carlo Markov Chain (MCMC) fitting was used to explore the amplitude-phase parameter space. The values inferred from the relations are therefore used only as a starting point for the MCMC walkers. The amplitude-phase combination that best scales the template to the observations is then used to obtain the mean magnitude for the Cepheid.  
 
\begin{figure}[t!]
    \centering
    \includegraphics[width=10cm]{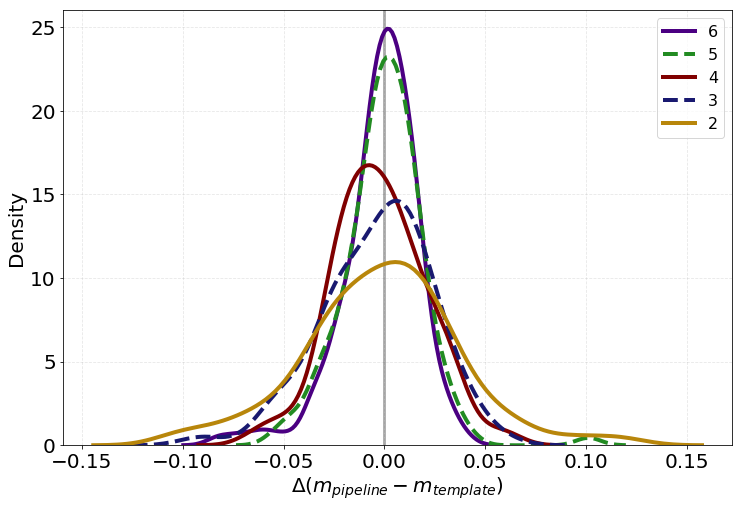}
    \caption{The distribution of the magnitude difference between magnitudes obtained from the photometry pipeline using all observations and the magnitudes obtained from template fitting using $N=2,3,4,5$ and $6$ observations. }
    \label{fig:mcmc}
\end{figure}

 To test how well the template fitting procedure performs on sparse data, we simulated data similar to the SAGE dataset by randomly removing observations for the calibrating sample, leaving between two and six observations for each Cepheid. We then ran the MCMC template fitting code on this data and compared the resulting mean magnitudes to those obtained from the photometry pipeline. Figure \ref{fig:mcmc} shows the distribution of magnitude differences between the two methods for $N=2,3,4,5$ and $6$ observations. The figure shows good agreement between both methods, particularly as the number of observations increases. 

{}

\section{Future work}

The next steps in this project are the application of the template fitting method to the SAGE data to obtain mid-infrared mean magnitudes, amplitudes, colours and light curves for all known fundamental mode classical Cepheids in the Magellanic Clouds. A calibrated LL from this data will then be used to obtain distances to individual Cepheids and construct a 3D map of the  Magellanic System, which will then be used to probe the geometric and chemical structure of the system. After this stage of the project is complete, the same process will be carried out for the RR Lyrae stars in the Magellanic Clouds. The 3D distribution of the young Cepheids will then be compared to the distribution of old RR Lyraes to study the dynamical history of the system.

\acknowledgements The authors would like to thank Raoul \& Catherine Hughes for awarding the RCH Alan Hunter Scholarship to fund this research. In addition, AC would like to thank the Santander Postgraduate Mobility Award and the IOP CR Barber Trust for providing funding to attend this conference. This work is based on observations made with the \textit{Spitzer Space Telescope}, which is operated by the Jet Propulsion Laboratory, California Institute of Technology under a contract with NASA.

\bibliographystyle{mnras}
\bibliography{library}  

\end{document}